\documentclass[10pt,conference]{IEEEtran}
\IEEEoverridecommandlockouts
\usepackage{amsmath,amsfonts,amssymb}
\usepackage{algorithmic}
\usepackage{algorithm}
\usepackage{array}
\usepackage{cite}
\usepackage{textcomp}
\usepackage{stfloats}
\usepackage{url}
\usepackage{verbatim}
\usepackage{graphicx}
\usepackage{siunitx}
\usepackage{bm}
\usepackage{etoolbox} 
\usepackage{xcolor}
\usepackage{microtype} 
\usepackage{caption}
\usepackage{subcaption} 
\usepackage{mathtools}
\usepackage{lipsum} 
\newtheorem{proposition}{\underline{Proposition}}
\allowdisplaybreaks

\begin{document}
	
	\title{\parbox{1\textwidth}{\centering Beyond Diagonal IRS Aided OFDM: Rate Maximization under Frequency-Dependent Reflection\\}}
	
	\author{
		{\large Ye Yuan and Shuowen Zhang}\\
		{\normalsize Department of Electrical and Electronic Engineering, The Hong Kong Polytechnic University}\\ 
		{\normalsize E-mails: joseph.yuan@connect.polyu.hk, shuowen.zhang@polyu.edu.hk}\\
		\vspace{-1cm}
	}

	\maketitle
	
	\pagestyle{empty}  
	\thispagestyle{empty}

	\begin{abstract}
		This paper studies a broadband orthogonal frequency division multiplexing (OFDM) system aided by a beyond diagonal intelligent reflecting surface (BD-IRS), where inter-connections exist among different elements such that the reflection matrix can exhibit a beyond diagonal structure. Under practical circuit structures, the reflection matrix of the BD-IRS is generally dependent on the circuit parameters (e.g., capacitance matrix for all tunable capacitors) as well as the operating frequency, which leads to couplings among the BD-IRS reflection matrices over different sub-carriers and consequently new challenges in the BD-IRS design. Motivated by this, we first model the relationship between the BD-IRS reflection matrices over different sub-carriers and the tunable capacitance matrix, and then formulate the joint optimization problem of the tunable capacitance matrix and power allocation over OFDM sub-carriers to maximize the achievable rate of the OFDM system. Despite the non-convexity of the problem, we propose an effective algorithm for finding a high-quality feasible solution via leveraging alternating optimization and successive convex approximation. Numerical results show the superiority of our proposed design over benchmark designs.
	\end{abstract}
	
	\vspace{-0.25cm}
	\section{Introduction}
\begin{figure}[t]
		\centering
		\includegraphics[width=0.36\textwidth]{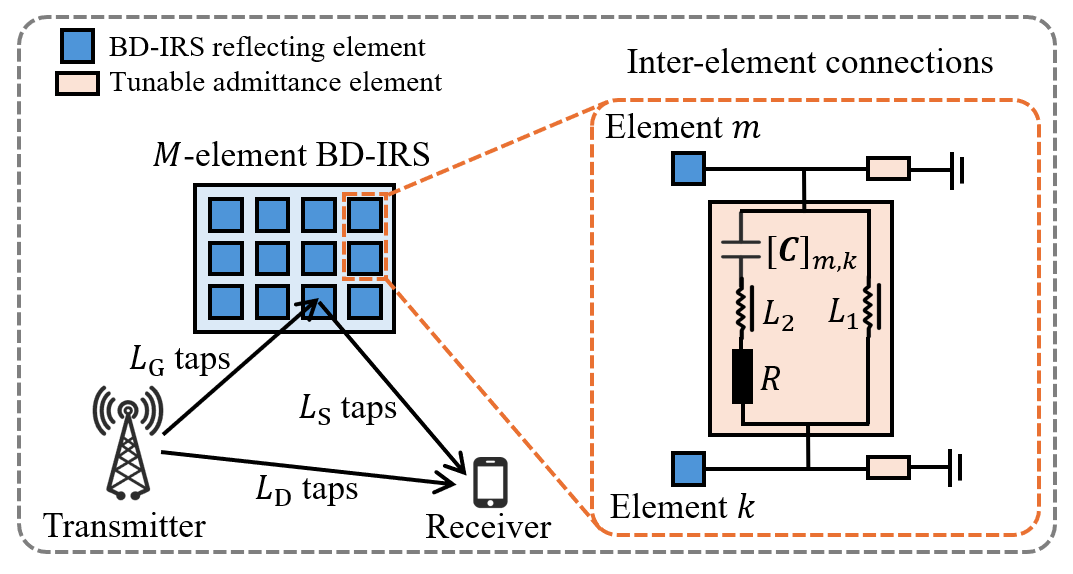} 
		\caption{Illustration of a BD-IRS aided broadband OFDM communication system.}
		\label{OFDM}
		\vspace{-0.85cm}
	\end{figure}

	\vspace{-0.15cm}
	\looseness=-1
	As a revolutionary technology, intelligent reflecting surface (IRS) employs passive reflecting elements to alter the propagation environment, thereby enhancing signal coverage and improving communication performance. In a conventional IRS, reflecting elements are connected solely to their self-admittance, which leads to a diagonal structure of the reflection matrix. To enhance reflection efficiency, the concept of beyond diagonal IRS (BD-IRS) has been introduced, inspired by multi-port network theory \cite{Shen2022Modeling}. As illustrated in Fig.~\ref{OFDM}, adding additional tunable admittance elements enables inter-connections among reflecting elements, leading to a reflection matrix that includes off-diagonal elements. This enhancement improves the flexibility of electromagnetic wave manipulation and has the potential to further increase the communication rate.  
	
	The performance improvement brought by BD-IRS in narrowband communication has been widely verified \cite{11054049}. However, its application to \emph{broadband} communication has received limited attention. On the other hand, although existing studies have explored the application of conventional diagonal IRS in broadband communication and proposed relevant designs \cite{Yang2020Intelligent, Yang2020IRS, Zhang2020Capacity, Li2021Intelligent}, these studies are difficult to be directly applied to BD-IRS due to the new reflection matrix structure. Specifically, the inter-connections among the BD-IRS elements make it impossible to model the phase and amplitude response of each element separately as in conventional IRS. Moreover, existing schemes mostly assumed that BD-IRS is \emph{frequency-independent}, while the complex circuit structure of the BD-IRS introduces \emph{frequency dependence}. Therefore, the effective modeling and optimization of practical frequency-dependent BD-IRS remains a challenge.

	Among the few related studies, \cite{Demir2024Wideband} and \cite{Soleymani2024Maximizing} explored the BD-IRS reflection design in broadband communication. However, they assumed that the BD-IRS is frequency-independent, which may not be accurate in practical broadband systems. On the other hand, \cite{Katsanos2024Multi} and \cite{Sena2024Beyond} studied the distributed total rate maximization and spectrum efficiency maximization problems via configuring one or more BD-IRS in multi-band multi-transmitter systems, respectively. \looseness=-1 However, frequency dependence was still not explicitly modeled and dealt with, while a priority frequency or switch array based method was adopted instead. 
	
	A recent work \cite{Li2024Beyond} attempted to optimize the frequency-dependent BD-IRS using a linear fitting method due to the inherent complexity. Specifically, \cite{Li2024Beyond} demonstrated that the frequency response of the BD-IRS can be approximated via a linear fitting model under a specific system with a central frequency of $2.4$ GHz and a bandwidth of $300$ MHz. However, this method cannot be directly generalized to broader frequency bands or more diverse systems to the best of our knowledge. Moreover, the fitting method may exhibit certain performance limit. How to systematically model and optimize the frequency response of BD-IRS for general systems still remains a challenging unaddressed problem, which motivates our study in this work.
	
	In this paper, we first provide an explicit modeling of the BD-IRS reflection matrices over different OFDM sub-carriers with respect to the tunable capacitance matrix for the BD-IRS circuit. Then, we study the joint optimization problem of the capacitance matrix and the power allocations over different OFDM sub-carriers, for the purpose of maximizing the OFDM achievable rate. This problem is challenging due to its non-convexity and that the BD-IRS reflection matrices over different sub-carriers are \emph{coupled} by the \emph{common} capacitance matrix for the BD-IRS circuit. To resolve these difficulties, we propose an alternating optimization (AO) based algorithm to find a high-quality suboptimal solution. Numerical results show that the proposed design outperforms various benchmarks.

	\vspace{-0.15cm}
	\section{System Model}
	\vspace{-0.05cm}
	We consider a broadband OFDM communication system with $ N \geq 1 $ sub-carriers denoted by set $\mathcal{N} = \{ 1, \ldots, N \}$. As illustrated in Fig.~\ref{OFDM}, a BD-IRS with $M \geq 1$ reflecting elements denoted by set $\mathcal{M} = \{ 1, \ldots, M \}$ is employed to enhance the communication from a single-antenna transmitter to a single-antenna receiver. 
	Let $L_{\mathrm{D}}$, $L_{\mathrm{G}}$, and $L_{\mathrm{S}}$ denote the numbers of delayed taps in the time-domain impulse responses for the direct link, the transmitter-IRS link, and the IRS-receiver link, respectively. Let $\bar{d}_{l}\in \mathbb{C}, \ \forall l \in \{0, \ldots, L_{\mathrm{D}}-1\}$, $ \bar{\bm{g}}_l \in \mathbb{C}^{M \times 1}, \ \forall l \in \{0, \ldots, L_{\mathrm{G}}-1\}$, and  $\bar{\bm{s}}_l^H\in \mathbb{C}^{1\times M}, \ \forall l \in \{0, \ldots, L_{\mathrm{S}}-1\}$ denote the corresponding time-domain channels at each $l$-th tap, respectively.
	The overall impulse response of the reflected link is the convolution of $\{\bar{\bm{s}}_l^H\}_{l=0}^{L_\mathrm{S}-1}$, the time-domain response of the BD-IRS, and $\{ \bar{\bm{g}}_l\}_{l=0}^{L_\mathrm{G}-1}$. Thus, the overall impulse response from the transmitter to the receiver consists of at most $L_\mathrm{max} = \max\{L_\mathrm{D}, L_\mathrm{G}+L_\mathrm{S}-1\}$ delayed taps.
	In the frequency domain, the channels for the direct link, the transmitter-IRS link, and the IRS-receiver link at each $n$-th sub-carrier are given by $d_n = \sum_{l=0}^{L_\mathrm{D}-1} \bar{d}_le^{-j2\pi(n-1)l/N}$, $\bm{g}_n = \sum_{l=0}^{L_\mathrm{G}-1} \bar{\bm{g}}_le^{-j2\pi(n-1)l/N}$, and $\bm{s}^H_n = \sum_{l=0}^{L_\mathrm{S}-1} \bar{\bm{s}}^H_le^{-j2\pi(n-1)l/N}$, respectively. 
	
	We consider a fully-connected BD-IRS that only performs reflection, and a general frequency response model of the BD-IRS which incorporates potential frequency-dependent reflection. Let $\boldsymbol{\Phi}_n \in \mathbb{C}^{M \times M}$ denote the BD-IRS reflection matrix at each $n$-th sub-carrier.
	Based on the convolution theorem, the overall effective frequency-domain channel from the transmitter to the receiver at each $n$-th sub-carrier is given by  $h_n = d_n + \bm{s}^H_n \boldsymbol{\Phi}_n \bm{g}_n$.
	We assume that perfect channel state information (CSI) is available at both the transmitter and the receiver. The achievable rate in bits per second per Hertz (bps/Hz) is then given by
	\vspace{-0.2cm}
	\begin{equation}
		r \!=\! \frac{1}{N+N_\mathrm{CP}}\sum_{n=1}^{N} \log_2 \left( 1 + \frac{p_n|d_n + \bm{s}^H_n \boldsymbol{\Phi}_n \bm{g}_n|^2}{\Gamma \sigma^2} \right), \label{r}
		\vspace{-0.2cm}
	\end{equation}
	where $N_\mathrm{CP} \geq L_\mathrm{max}$ denotes the cyclic prefix (CP) length; $p_n$ denotes the power allocation at each $n$-th sub-carrier under \( p_n \geq 0 \) and \( \sum_{n=1}^{N} p_n \leq P \), with $P$ denoting the total transmission power; $\Gamma \geq 1$ denotes the achievable rate gap due to practical modulation and coding scheme (MCS) \cite{CioffiMulti}; $\sigma^2$ denotes the average receiver noise power per sub-carrier.

	\vspace{-0.1cm}
	\section{Frequency Response Model of BD-IRS}
	\vspace{-0.1cm}
	In this section, we model the frequency response of the BD-IRS, i.e., the reflection matrices over different frequencies. As shown in Fig.~\ref{OFDM}, BD-IRS comprises inter-element connections with intricate admittance circuit structure, based on which we model the frequency response of BD-IRS as follows. Specifically, we consider a lumped circuit as illustrated in Fig.~\ref{OFDM} for each tunable admittance between the $m$-th element and the $k$-th element, which consists of one fixed resistor with resistance $R$~\text{Ohm ($\Omega$)}, two fixed inductors each with inductance $L_1$ or $L_2$~\text{Henry (H)}, and one tunable capacitor with \emph{tunable capacitance} \( C_{m,k} \geq 0\)~\text{Farad (F)}. Similarly, the admittance between each $m$-th element and the ground consists of one fixed resistor with resistance $R$~$\Omega$, two fixed inductors each with inductance $L_1$ or $L_2$~H, and one tunable capacitor with \emph{tunable capacitance} \( C_{m,m} \geq 0\)~F. The overall tunable capacitance matrix for the BD-IRS is denoted by $\bm{C} \in \mathbb{R}^{M \times M}$, where $[\bm{C}]_{m,k} = C_{m,k}, \ \forall m,k \in \mathcal{M}$.
	
	\looseness=-1
	Note that the admittance between each two elements and that between each element and the ground are determined by the aforementioned resistance, inductance, and capacitance, as well as the operating frequency \cite{Li2024Beyond}. With a large number of OFDM sub-carriers, the sub-carrier spacing is generally small, thus the admittance over each sub-carrier is approximately the same.
	Let $f_\mathrm{c}$ and $B$ denote the central frequency of the OFDM system and the system bandwidth, respectively. The frequency of each $n$-th sub-carrier is given by $f_n = f_\mathrm{c} + \frac{B}{N}(n-\frac{N+1}{2})$.
	Let \( \bm{A}_n \in \mathbb{C}^{M \times M}\) denote the admittance matrix at each \( n \)-th sub-carrier, in which each entry can be shown to be given by \cite{Nerini2024A}
	\vspace{-0.25cm}
	\setlength{\belowdisplayskip}{3pt}
	\begin{align}
		&[\bm{A}_n]_{m,k} = \notag \\
		&\begin{cases}
			\begin{aligned}
				&  \sum\limits_{\bar{k}=1}^{M} \left(\frac{1}{R + j 2 \pi f_n L_2 \!+\! \frac{1}{j 2 \pi f_n [\bm{C}]_{m,\bar{k}}}} \!+\! \frac{1}{j 2 \pi f_n L_1} \right),  m = k, 
			\end{aligned} \\
			\begin{aligned}
				& -\left(\frac{1}{R + j 2 \pi f_n L_2 \!+\! \frac{1}{j 2 \pi f_n [\bm{C}]_{m,k}}} \!+\! \frac{1}{j 2 \pi f_n L_1}\right),  m\neq k.
			\end{aligned}
		\end{cases} \label{admittance}
	\end{align}
	
	According to network theory \cite{Pozar2009Microwave}, the BD-IRS reflection matrix at each $n$-th sub-carrier is determined by its corresponding admittance matrix as
	\vspace{-0.15cm}
	\setlength{\belowdisplayskip}{3pt}
	\begin{equation}
		\boldsymbol{\Phi}_n = (a_0 \bm{I}_M + \bm{A}_n)^{-1}(a_0 \bm{I}_M - \bm{A}_n), \ \forall n \in \mathcal{N},\label{Phi_from_A}
		\vspace{-0.05cm}
	\end{equation}
	where $a_0 > 0$ denotes the characteristic admittance. Note that the admittance circuit structure can be either symmetric with an additional constraint of $\bm{C} = \bm{C}^T$ and consequently $\boldsymbol{\Phi}_n = \boldsymbol{\Phi}_n^T$, or general without such symmetric constraint \cite{10993452}. In this paper, we focus on the general structure without the symmetric constraint, while our results can be directly extended to the case with the additional symmetric constraint, which will be evaluated numerically in Section~\ref{Numerical_results}.
	
	Note that $\boldsymbol{\Phi}_n$ is determined by both the sub-carrier frequency $f_n$ and the circuit parameters, in which the only tunable component is the overall capacitance matrix $\bm{C}$. This thus motivates our optimization of $\bm{C}$ such that the BD-IRS response and consequently the effective channel is most favorable for broadband OFDM communication. Moreover, it is worth noting that the reflection matrices over different sub-carriers are \emph{coupled} by a \emph{common} capacitance matrix $\bm{C}$, thus the design of $\bm{C}$ should cater to the frequency-domain channels $\bm{s}_n^H$'s, $\bm{g}_n$'s, and $d_n$'s for all sub-carriers, which makes the problem particularly challenging.

	\vspace{-0.1cm}
	\section{Problem Formulation}
	\vspace{-0.1cm}
	\looseness=-1
	In this section, we formulate the problem of jointly optimizing the tunable capacitance matrix $\bm{C}$ and the sub-carrier power allocation \(\{p_n\}_{n=1}^N\) to maximize the achievable rate of the BD-IRS aided OFDM system. By dropping the constant term $\frac{1}{N+N_{\mathrm{CP}}}$ in \eqref{r}, the optimization problem is formulated as
	\vspace{-0.25cm}
	\setlength{\belowdisplayskip}{3pt}
	\begin{align}
		\mbox{(P1)}   \ \ \ \ \notag \\
		\mathop{\rm{max}}_{\bm{C}, \{p_n\}_{n=1}^N, \atop {\{\boldsymbol{\Phi}_n\}_{n=1}^N, \atop \{\bm{A}_n\}_{n=1}^N}}
		& \sum_{n=1}^{N} \log_2 \left( 1 + \frac{p_n|d_n + \bm{s}^H_n \boldsymbol{\Phi}_n \bm{g}_n|^2 }{\Gamma \sigma^2} \right)  \label{objective_function} \\
		\mathrm{s.t.} \ \ \ \,
		& \! \boldsymbol{\Phi}_n  \!=\!  ( a_0 \bm{I}_M  + \! \bm{A}_n  )^{-1} ( a_0 \bm{I}_M  -  \bm{A}_n  ), \ \forall  n \! \in \! \mathcal{N} \label{Phi_A} \\
		& \! [\bm{A}_n]_{m,k} = \notag \\
		&  \!  \begin{cases}
			\begin{aligned}
				&  \sum\limits_{\bar{k}=1}^{M} \Big(\frac{1}{R \!+\! j 2 \pi f_n L_2 \!+\! \frac{1}{j 2 \pi f_n [\bm{C}]_{m,\bar{k}}}} \!+\! \frac{1}{j 2 \pi f_n L_1} \Big), \\
				& \ m = k, 
			\end{aligned} \\
			\begin{aligned}
				& -\Big(\frac{1}{R \!+\! j 2 \pi f_n L_2 \!+\! \frac{1}{j 2 \pi f_n [\bm{C}]_{m,k}}} \!+\! \frac{1}{j 2 \pi f_n L_1}\Big), \\
				&  \  m\neq k,
			\end{aligned}
		\end{cases} \notag \\
		& \ \forall m, k \in \mathcal{M}, \ \forall n \in \mathcal{N} \label{constraint_2}\\ 
		&  [\bm{C}]_{m,k} \geq 0, \ \forall m,k \in \mathcal{M} \label{constraint_C}\\     
		&  \sum_{n=1}^{N} p_n \leq P  \label{constraint_4} \\ 
		&  p_n \geq 0, \ \forall n \in \mathcal{N}.\label{constraint_5}
		\vspace{-0.1cm}
	\end{align}
	(P1) is a non-convex optimization problem because the objective function is non-concave over $\{\boldsymbol{\Phi}_n\}_{n=1}^N$ and $\{ p_n \}_{n=1}^N$, and the constraints in \eqref{Phi_A} and \eqref{constraint_2} can be shown to be non-convex. Moreover, $\{\boldsymbol{\Phi}_n\}_{n=1}^N$ and $\{ p_n \}_{n=1}^N$ are coupled in the objective function, and the relationship between $\{\boldsymbol{\Phi}_n\}_{n=1}^N$ and $\bm{C}$ via $\{\bm{A}_n\}_{n=1}^N$ is highly complex, which make (P1) more difficult to solve. In the following, we propose a relaxation-based approach to deal with the complex expression of $\{\boldsymbol{\Phi}_n\}_{n=1}^N$ with respect to $\bm{C}$, and develop an AO-based algorithm to iteratively optimize each of the coupled variables.
	
	\vspace{-0.1cm}
	\section{Proposed Solution to (P1)}
	\addtolength{\topmargin}{0.02in}
	\vspace{-0.1cm}
	\subsection{Relaxation of (P1)}
	Firstly, we replace the constraints in \eqref{Phi_A} and \eqref{constraint_2} with a set of relaxed constraints on $\{\boldsymbol{\Phi}_n\}_{n=1}^N$. To this end, we present the following proposition.
	
	\begin{proposition} \label{proposition1}
		Any \(\{\boldsymbol{\Phi}_n\}_{n=1}^N \) under \eqref{Phi_from_A} and \eqref{admittance} is guaranteed to satisfy $\boldsymbol{\Phi}_n \boldsymbol{\Phi}_n^H \preceq \bm{I}_M, \ \forall n \in \mathcal{N}$.
	\end{proposition}
	
	\textit{Proof:} 
	Let $\bm{X}_n = a_0\bm{I}_M + \bm{A}_n$ and $\bm{Y}_n = a_0\bm{I}_M - \bm{A}_n$. Then
	\begin{align}
		\bm{I}_M -\boldsymbol{\Phi}_n\boldsymbol{\Phi}_n^H &=\bm{X}_n^{-1}\big(\bm{X}_n\bm{X}_n^H-\bm{Y}_n\bm{Y}_n^H\big)(\bm{X}_n^{-1})^H \notag \\
		&=2a_0\bm{X}_n^{-1}(\bm{A}_n+\bm{A}_n^H)(\bm{X}_n^{-1})^H.
	\end{align}
	Note that $\bm{A}_n$ is the admittance matrix of a passive network, which satisfies $\bm{A}_n + \bm{A}_n^H \succeq \bm{0}$ \cite{Pozar2009Microwave}.
	Since $a_0>0$ and $\bm{X}_n$ is invertible, $\bm{I}_M -\boldsymbol{\Phi}_n\boldsymbol{\Phi}_n^H\succeq\bm{0}$ holds, i.e., $\boldsymbol{\Phi}_n \boldsymbol{\Phi}_n^H \preceq \bm{I}_M$. This completes the proof of Proposition~\ref{proposition1}.
	
	Based on the results of Proposition~\ref{proposition1}, we transform (P1) into the following relaxed problem, with the constraints in \eqref{Phi_A} and \eqref{constraint_2} replaced with the relaxed constraints in \eqref{constraint_6}:
	\vspace{-0.25cm}
	\setlength{\belowdisplayskip}{3pt}
	\begin{align}
		\mbox{(P2)} \ \mathop{\rm{max}}_{ \{p_n\}_{n=1}^N, \atop \{\boldsymbol{\Phi}_n\}_{n=1}^N} \
		& \sum_{n=1}^{N} \log_2 \left( 1 + \frac{p_n|d_n + \bm{s}^H_n \boldsymbol{\Phi}_n \bm{g}_n|^2}{\Gamma \sigma^2} \right) \\
		\mathrm{s.t.} \ \ \,     
		& \boldsymbol{\Phi}_n \boldsymbol{\Phi}_n^H \preceq \bm{I}_M,  \ \forall n \in \mathcal{N} \label{constraint_6} \\
		& \eqref{constraint_4},\eqref{constraint_5}.  \notag
		\vspace{-0.1cm}
	\end{align}
	Note that although the new constraints in \eqref{constraint_6} are convex, (P2) is still non-convex due to the coupling between $\{\boldsymbol{\Phi}_n\}_{n=1}^N$ and $\{ p_n \}_{n=1}^N$ in the objective function. In the following, we leverage the AO method to iteratively optimize $\{\boldsymbol{\Phi}_n\}_{n=1}^N$ and $\{ p_n \}_{n=1}^N$, respectively.

	\vspace{-0.1cm}
	\subsection{AO Method for (P2)} \label{AO_Method}
	\vspace{-0cm}
	In the AO method, we start with a given initial point of $\{\boldsymbol{\Phi}_n\}_{n=1}^N$, and iteratively optimize $\{ p_n \}_{n=1}^N$ and $\{\boldsymbol{\Phi}_n\}_{n=1}^N$ with the other being fixed at each time. Specifically, with given $\{\boldsymbol{\Phi}_n\}_{n=1}^N$, (P2) is reduced to the following sub-problem:
	\vspace{-0.45cm}
	\setlength{\belowdisplayskip}{3pt}
	\begin{align}
		\mbox{(P2-I)} \ \mathop{\rm{max}}_{ \{p_n\}_{n=1}^N} \ 
		& \sum_{n=1}^{N} \log_2 \left( 1 + \frac{p_n|d_n + \bm{s}^H_n \boldsymbol{\Phi}_n \bm{g}_n|^2 }{\Gamma \sigma^2} \right) \\
		\mathrm{s.t.} \ \ \   
		& \eqref{constraint_4}, \eqref{constraint_5}. \notag
		\vspace{-0.2cm}
	\end{align}
	(P2-I) can be solved by the water-filling (WF) algorithm \cite{goldsmith2005wireless}, for which the optimal solution is given by $p_n^\star = \left( \frac{1}{\lambda} - \frac{\Gamma \sigma^2}{|d_n + \bm{s}^H_n \boldsymbol{\Phi}_n \bm{g}_n|^2}\right)^+$, with $(x)^+ \triangleq \max\{x,0\}$ and $\lambda$ denoting the cut-off channel-to-noise power ratio that satisfies $\sum_{n=1}^{N} \left( \frac{1}{\lambda} - \frac{\Gamma \sigma^2}{|d_n + \bm{s}^H_n \boldsymbol{\Phi}_n \bm{g}_n|^2} \right)^+ = P$.
	On the other hand, with given $\{p_n\}_{n=1}^N$, (P2) is reduced to the following sub-problem:
	\vspace{-0.45cm}
	\setlength{\belowdisplayskip}{3pt}
	\begin{align}
		\mbox{(P2-II)} \ \mathop{\rm{max}}_{ \{\boldsymbol{\Phi}_n\}_{n=1}^N} \
		& \sum_{n=1}^{N} \log_2 \left( 1 + \frac{p_n|d_n + \bm{s}^H_n \boldsymbol{\Phi}_n \bm{g}_n|^2 }{\Gamma \sigma^2}\right) \\
		\mathrm{s.t.} \ \ \    
		& \boldsymbol{\Phi}_n \boldsymbol{\Phi}_n^H \preceq \bm{I}_M,  \ \forall n \in \mathcal{N}.
	\end{align}
	Note that (P2-II) is still non-convex due to the non-concave objective function with respect to $\{\boldsymbol{\Phi}_n\}_{n=1}^N$. To resolve this issue, we propose to apply the successive convex approximation (SCA) technique \cite{marks1978A} to successively approximate the objective function with its concave lower bound. Specifically, by introducing auxiliary real variables $y_n$, $a_n$,  and $b_n$, (P2-II) can be equivalently transformed as the following problem:
	\vspace{-0.25cm}
	\setlength{\belowdisplayskip}{3pt}
	\begin{align}
		\mbox{(P2-II-\text{eqv})} \! \mathop{\rm{max}}_{ \{\boldsymbol{\Phi}_n\}_{n=1}^N, \atop {\{y_n\}_{n=1}^N, \atop {\{a_n\}_{n=1}^N, \atop \{b_n\}_{n=1}^N}}} 
		& \sum_{n=1}^{N} \log_2 \left( 1 + \frac{p_n y_n }{\Gamma \sigma^2}\right) \\
		\mathrm{s.t.}  \ \ \,
		& \boldsymbol{\Phi}_n \boldsymbol{\Phi}_n^H \preceq \bm{I}_M,  \ \forall n \in \mathcal{N} \label{passive} \\
		& a_n \!=\! \mathfrak{Re} \{ d_n \!+\! \bm{s}^H_n \boldsymbol{\Phi}_n \bm{g}_n \}, \forall n \in \mathcal{N} \label{a_n} \\
		& b_n \!=\! \mathfrak{Im} \{ d_n \!+\! \bm{s}^H_n \boldsymbol{\Phi}_n \bm{g}_n \}, \forall n \in \mathcal{N} \label{b_n}\\
		& y_n \leq \tilde{g}_n(a_n, b_n), \ \forall n \in \mathcal{N},
	\end{align}
	where $\mathfrak{Re}\{\cdot\}$ and $\mathfrak{Im}\{\cdot\}$ denote the real and imaginary parts of a complex number, respectively; $\tilde{g}_n(a_n, b_n) \triangleq a_n^2 + b_n^2$ is a convex and differentiable function over $a_n$ and $b_n$.
	At any given local point $(\tilde{a}_n, \tilde{b}_n)$, the first-order approximation of $\tilde{g}_n(a_n, b_n)$ given by $g_n(a_n, b_n)$ below is a concave lower bound of $\tilde{g}_n(a_n, b_n)$:
	\vspace{-0.3cm}
	\begin{equation}
		\tilde{g}_n(a_n, b_n)  \! \geq \! \tilde{a}_n^2 \!+\! \tilde{b}_n^2 + \begin{bmatrix} 2 \tilde{a}_n \\ 2 \tilde{b}_n \end{bmatrix}^T \! \begin{bmatrix} a_n - \tilde{a}_n \\ b_n - \tilde{b}_n \end{bmatrix} \! \triangleq \! g_n(a_n, b_n) \label{f2},
		\vspace{-0.15cm}
	\end{equation}
	where the equality holds if and only if $a_n = \tilde{a}_n$ and $b_n = \tilde{b}_n$. Note that $g_n(a_n, b_n)$ has the same gradient as $\tilde{g}_n(a_n, b_n)$ at $(\tilde{a}_n, \tilde{b}_n)$. Then, an approximate solution to (P2-II-\text{eqv}) and (P2-II) can be obtained by iteratively solving the following problem with the local points $(\tilde{a}_n, \tilde{b}_n)$ updated as the optimal solution in the previous iteration:
	\vspace{-0.2cm}
	\setlength{\belowdisplayskip}{3pt}
	\begin{align}
		\!\!\! \mbox{(P2-\text{II-eqv}')}  \mathop{\rm{max}}_{ \{\boldsymbol{\Phi}_n\}_{n=1}^N, \{y_n\}_{n=1}^N, \atop \{a_n\}_{n=1}^N, \{b_n\}_{n=1}^N} 
		& \sum_{n=1}^{N} \log_2 \left( 1 + \frac{p_n y_n}{\Gamma \sigma^2} \right) \\
		\mathrm{s.t.} \quad \quad \
		& \eqref{passive}, \eqref{a_n}, \eqref{b_n} \notag \\
		& y_n \leq g_n(a_n, b_n), \ \forall n \in \mathcal{N}.
		\vspace{-0.2cm}
	\end{align}
	Note that (P2-II-eqv') is a convex optimization problem which can be solved via the interior-point method or CVX \cite{cvx}.
	It can be shown that the above update process guarantees monotonic convergence and is guaranteed to reach at least a stationary point of (P2-II) \cite{marks1978A}.

	To summarize, by iteratively updating $\{{p_n}\}_{n=1}^N$ and $\{\boldsymbol{\Phi}_n\}_{n=1}^N$ via obtaining the optimal solution to (P2-I) and the stationary point of (P2-II), respectively, the AO-based algorithm is guaranteed to converge monotonically to at least a stationary point of (P2) \cite{Hong2016Unified}. It is worth noting that the obtained solution to (P2) may not satisfy the original constraints for the BD-IRS reflection matrices under the circuit structure, i.e., the constraints in \eqref{Phi_A} and \eqref{constraint_2} of (P1). In the following, we construct a feasible solution to (P1) by designing the tunable capacitance matrix $\bm{C}$ based on the obtained $\{\boldsymbol{\Phi}_n\}_{n=1}^N$.	
	
	\vspace{-0.1cm}
	\subsection{Construction of Feasible Solution to (P1)}
	Note that with given BD-IRS reflection matrices $\{\boldsymbol{\Phi}_n\}_{n=1}^N$, the admittance matrix can be derived below based on \eqref{Phi_from_A}:
	\vspace{-0.2cm}
	\begin{equation}
		\bm{A}_n = a_0( \bm{I}_M - \boldsymbol{\Phi}_n)(\bm{I}_M + \boldsymbol{\Phi}_n)^{-1}, \ \forall n \in \mathcal{N}. \label{A_from_Phi}
		\vspace{-0.1cm}
	\end{equation}
	Based on \eqref{admittance}, the tunable capacitance matrix that corresponds to $\bm{A}_n$ can be derived as
	\vspace{-0.2cm}
	\setlength{\belowdisplayskip}{3pt}
	\begin{align}
		&[\bm{C}]_{m,k} = [\tilde{\bm{C}}_n]_{m,k} \overset{\Delta}{=} \notag \\
		&\begin{cases}
			\begin{aligned}
				&  \frac{1}{j 2 \pi f_n  \left( \frac{1}{\sum_{\bar{k}=1}^{M}[\bm{A}_n]_{m,\bar{k}} - \frac{1}{j 2 \pi f_n  L_1}} - R - j 2 \pi f_n  L_2 \right)},  m = k,
			\end{aligned} \\
			\begin{aligned}
				& \frac{1}{j 2 \pi f_n  \left( \frac{1}{-[\bm{A}_n]_{m,k} - \frac{1}{j 2 \pi f_n  L_1}} - R - j 2 \pi f_n  L_2 \right)},  m\neq k.
			\end{aligned}
		\end{cases} \label{xxx}
	\end{align}
	However, note that $\tilde{\bm{C}}_n$'s in \eqref{xxx}, i.e., the favorable capacitance matrices for each of the $N$ sub-carriers, are generally different for different $n$'s, thus \eqref{xxx} generally cannot hold for all $n \in \mathcal{N}$ due to the common $\bm{C}$ for the BD-IRS circuit. To recover a common capacitance matrix, we formulate the following optimization problem that aims to minimize the overall difference between $\bm{C}$ and $\tilde{\bm{C}}_n$'s:
	\vspace{-0.2cm}
	\setlength{\belowdisplayskip}{3pt}
	\begin{align}
		\mbox{(P3)} \ \mathop{\rm{min}}_{\bm{C}} \
		& \sum_{n=1}^{N} \|\bm{C} - \tilde{\bm{C}}_n \|_F \\
		\mathrm{s.t.} \      
		& [\bm{C}]_{m,k} \geq 0, \ \forall m,k \in \mathcal{M},
	\end{align}
	where $\| \cdot \|_F$ denotes the Frobenius norm. The objective of (P3) is to find a common $\bm{C}$ to ``represent'' the favorable capacitance matrices for all sub-carriers. (P3) is a convex problem, which can be solved via CVX \cite{cvx}. Based on the obtained $\bm{C}$, we will solve (P2-I) to obtain the corresponding optimal $\{p_n\}_{n=1}^N$.

	\vspace{-0.1cm}
	\subsection{Initialization of the Proposed Algorithm}
	\vspace{-0cm}
	The proposed AO-based algorithm and consequently the quality of the constructed feasible solution to (P1) are critically dependent on the choice of the initial point. In this subsection, we propose an initialization method for the AO-based algorithm. Specifically, we aim to maximize the total effective channel power over all OFDM sub-carriers via solving the following problem:
	\vspace{-0.25cm}
	\setlength{\belowdisplayskip}{3pt}
	\begin{align}
		\mbox{(P4)} \ \mathop{\rm{max}}_{\{\boldsymbol{\Phi}_n\}_{n=1}^N} \ 
		& \sum_{n=1}^{N} |d_n + \bm{s}^H_n \boldsymbol{\Phi}_n \bm{g}_n|^2 \\
		\mathrm{s.t.} \ \ \
		& \boldsymbol{\Phi}_n \boldsymbol{\Phi}_n^H \preceq \bm{I}_M,  \ \forall n \in \mathcal{N} \label{passive_1}. 
	\end{align}
	Note that although (P4) has a simpler objective function than the original rate function, it is still non-convex. To deal with it, we first transform it into a more tractable form by vectorizing each $\boldsymbol{\Phi}_n$. Based on the Kronecker identity \cite{Sena2024Beyond}, we have
	\vspace{-0.2cm}
	\begin{equation}
		\bm{s}^H_n \boldsymbol{\Phi}_n \bm{g}_n = \mathrm{vec}(\bm{s}^H_n \boldsymbol{\Phi}_n \bm{g}_n) = (\bm{g}_n^T \otimes \bm{s}^H_n)\mathrm{vec}(\boldsymbol{\Phi}_n),
		\vspace{-0.1cm}
	\end{equation}
	where $\mathrm{vec}(\cdot)$ and $\otimes$ denote the vectorization operation and the Kronecker product, respectively. Let $\bm{t}_n^H \triangleq \bm{g}_n^T \otimes \bm{s}^H_n \in \mathbb{C}^{1 \times M^2}$ and $\bm{q}_n \triangleq \mathrm{vec}(\boldsymbol{\Phi}_n) \in \mathbb{C}^{M^2 \times 1}$. Then, we have the following proposition.
	\begin{proposition} \label{proposition2}
		With $\bm{q}_n = \mathrm{vec}(\boldsymbol{\Phi}_n)$, $\boldsymbol{\Phi}_n \boldsymbol{\Phi}_n^H \preceq \bm{I}_M$ is equivalent to $\bm{q}_n^H(\bm{I}_M \otimes \bm{1}_M)\bm{q}_n \leq M$, where $\bm{1}_M$ denotes an $M \times M$ all-one matrix.
	\end{proposition}
	
	\textit{Proof:}
	With $\boldsymbol{\Phi}_n \boldsymbol{\Phi}_n^H \preceq \bm{I}_M$, it can be shown that $\bm{x}^H (\boldsymbol{\Phi}_n \boldsymbol{\Phi}_n^H - \bm{I}_M) \bm{x} \leq 0, \ \forall \bm{x} \in \mathbb{C}^{M \times 1}$. This implies $\|\boldsymbol{\Phi}_n^H \bm{x}\|^2 \leq \|\bm{x}\|^2$. 
	Since $\boldsymbol{\Phi}_n^H \bm{x} = \mathrm{vec}(\boldsymbol{\Phi}_n^H \bm{x}) = (\bm{x}^T \otimes \bm{I}_M) \mathrm{vec}(\boldsymbol{\Phi}_n^H)$, it follows that $\|\boldsymbol{\Phi}_n^H \bm{x}\|^2 = \|(\bm{x}^T \otimes \bm{I}_M) \mathrm{vec}(\boldsymbol{\Phi}_n^H)\|^2 = \|(\bm{I}_M \otimes \bm{x}^H) \bm{q}_n\|^2 \leq \|\bm{x}\|^2$. Consequently, $\bm{q}_n^H (\bm{I}_M \otimes \bm{x} \bm{x}^H) \bm{q}_n \leq \|\bm{x}\|^2, \ \forall \bm{x} \in \mathbb{C}^{M \times 1}$ holds. By setting $\bm{x}$ as a vector of all ones, it holds that $\bm{q}_n^H (\bm{I}_M \otimes \bm{1}_M) \bm{q}_n \leq M$. On the other hand, it can be shown that with $\bm{q}_n^H (\bm{I}_M \otimes \bm{1}_M) \bm{q}_n \leq M$, $\boldsymbol{\Phi}_n \boldsymbol{\Phi}_n^H \preceq \bm{I}_M$ is guaranteed to hold. 
	This thus completes the proof of Proposition~\ref{proposition2}.
	
	\addtolength{\topmargin}{0.01in}
	
	Thus, (P4) can be equivalently transformed as
	\vspace{-0.25cm}
	\setlength{\belowdisplayskip}{3pt}
	\begin{align}
		\mbox{(P4-eqv)} \! \mathop{\rm{max}}_{\{\bm{q}_n\}_{n=1}^N}  
		& \! \sum_{n=1}^{N}  \left( |d_n|^2 \! + \! d_n^\ast \bm{t}_n^H\bm{q}_n \! + \! d_n\bm{q}_n^H\bm{t}_n \! + \! |\bm{q}_n^H\bm{t}_n|^2 \right) \\
		\mathrm{s.t.} \ \
		& \bm{q}_n^H(\bm{I}_M \otimes \bm{1}_M)\bm{q}_n \leq M, \ \forall n \in \mathcal{N}.
	\end{align}
	Note that (P4-eqv) is a non-convex quadratically-constrained quadratic program (QCQP), for which we propose to apply the semi-definite relaxation (SDR) technique \cite{Luo2010Semidefinite}. Define
	\vspace{-0.2cm}
	\begin{align}
		\bm{M}_n &= \begin{bmatrix}
			\bm{t}_n\bm{t}_n^H & d_n\bm{I}_{M^2} \\
			d_n^\ast \bm{I}_{M^2} & \bm{0}
		\end{bmatrix}, \ \forall n \in \mathcal{N},\\
		\bm{D} &= \begin{bmatrix}
			\bm{I}_M \otimes \bm{1}_M & \bm{0} \\
			\bm{0} & \bm{0}
		\end{bmatrix}.
	\end{align}
	Let $\bm{w}_n \overset{\Delta}{=} [\bm{q}_n^T, \! \bm{t}_n^T]^T $, $\bm{W}_n \overset{\Delta}{=} \bm{w}_n\bm{w}_n^H, \ \forall n\in \mathcal{N}$. (P4-eqv) is equivalent to the following problem with additional constraints of  $\mathrm{rank}(\bm{W}_n) = 1, \forall n\in \mathcal{N}$:
	\vspace{-0.2cm}
	\begin{align}
		\mbox{(P4-SDR)} \ \mathop{\mathrm{max}}_{\{\bm{W}_n\}_{n=1}^N} \ 
		& \sum_{n=1}^{N} \Big( |d_n|^2 + \mathrm{tr}(\bm{W}_n\bm{M}_n) \Big) \\
		\text{s.t.} \ \ \ \,     
		& \mathrm{tr}(\bm{W}_n\bm{D}) \leq  M, \ \forall n \in \mathcal{N} \\
		& [\bm{W}_n]_{m,m} \!=\! \left| [\bm{t}_n]_{m-M^2} \right|^2, \ \forall m \in \notag \\
		&  \left\{M^2+1, \ldots, 2M^2 \right\}, \ \forall n \in \mathcal{N} \\
		& \bm{W}_n \succeq \bm{0}, \ \forall n \in \mathcal{N}. \label{W_n}
	\end{align}
	(P4-SDR) is a convex semi-definite program (SDP), which can be solved via CVX \cite{cvx}. Let $\{\bm{W}_n^\star\}_{n=1}^N$ denote the optimal solution to (P4-SDR).
	If $\mathrm{rank}(\bm{W}_n^\star)=1$ holds for all $n\in \mathcal{N}$, the SDR is tight, and (P4-SDR) is equivalent to (P4-eqv). The optimal solution to (P4-eqv) can be obtained as $\bm{q}_n = \bm{V}_n \mathrm{Diag}\{\boldsymbol{\Lambda}_n^{\frac{1}{2}}\}$,
	where $\bm{V}_n \boldsymbol{\Lambda}_n \bm{V}_n^H$ denotes the eigenvalue decomposition (EVD) of the upper-left $M^2 \times M^2$ submatrix of $\bm{W}_n^\star$; $\mathrm{Diag}\{\boldsymbol{\Lambda}_n^{\frac{1}{2}}\}$ extracts the main diagonal of $\boldsymbol{\Lambda}_n^{\frac{1}{2}}$ as a column vector. 
	On the other hand, if there exists one or more $n$'s such that $\mathrm{rank}(\bm{W}_n^\star)>1$, (P4-SDR) is not equivalent to (P4-eqv), and an approximate rank-one solution of $\bm{W}_n$ can be found via the Gaussian randomization method \cite{Zhang2018Constant} by generating $Q>1$ Gaussian random vectors under covariance matrix $\bm{W}_n$ and selecting the best one. Finally, the solution of each $\boldsymbol{\Phi}_n$ to (P4) can be obtained by expanding the obtained $\bm{q}_n$ for (P4-eqv) to the matrix form.
	
	\vspace{-0.1cm}
	\subsection{Complexity Analysis}
	\vspace{-0.1cm}
	The computational complexities for solving (P2-II-eqv'), (P3), and (P4-SDR) can be shown to be $\mathcal{O}(N^{3.5}M^9)$, $\mathcal{O}(NM^2)$, and $\mathcal{O}(NM^9)$, respectively. The worst-case complexity of recovering rank-one solutions of $\bm{W}_n$'s from the optimal solution to (P4-SDR) can be shown to be $\mathcal{O}(N M^6 + N Q M^4)$. Let $I_{\mathcal{SCA}}$ denote the worst-case number of iterations for the SCA method for (P2-II) to converge, and $I_{\mathcal{AO}}$ denote the number of times that the AO method iteratively updates $\{p_n\}_{n=1}^N$ and $\{\boldsymbol{\Phi}_n\}_{n=1}^N$. The worst-case overall complexity of the proposed algorithm can be shown to be $\mathcal{O}\big(N^{3.5}M^9I_{\mathcal{AO}}I_{\mathcal{SCA}} + NM^9 + N Q M^4 \big)$.
	\addtolength{\topmargin}{-0.02in}
	\vspace{-0.15cm}
	\section{Numerical Results} \label{Numerical_results}
	\vspace{-0.1cm}
	In this section, we evaluate the performance of the proposed algorithm via numerical results. We set $f_c = 2.4$~\text{GHz}, $B = 300$~\text{MHz}, $N = 64$, and $b= \frac{B}{N} = 4.6875$~\text{MHz}.
	For the number of delayed taps in each link, we set $L_{\mathrm{D}} = 16$, $L_{\mathrm{G}}=9$, and $L_{\mathrm{S}}=8$. The CP length is set as $N_\mathrm{CP}=16$.
	Let $\beta = \beta_0 d^{-\varepsilon}$ denote the distance-dependent path power model, with $\beta_0 = -30$~\text{dB} denoting the reference path power at reference distance $1$~\text{meters (m)}. Let $d_\mathrm{D} = 33$~\text{m}, $d_\mathrm{G}= 30$~\text{m}, and $d_\mathrm{S}= 5$~\text{m} denote the distance of the direct link, the transmitter-IRS link, and the IRS-receiver link, respectively. Let $\varepsilon_\mathrm{D} = 3.5$, $\varepsilon_\mathrm{G}= 2.2$, and $\varepsilon_\mathrm{S}= 2.8$ denote the corresponding path power exponents, respectively.
	Each time-domain channel coefficient is assumed to be an independent circularly symmetric complex Gaussian random variable with zero mean and variance following the exponential power delay profile, i.e., $d_l \sim \mathcal{CN}(0, \frac{\beta_\mathrm{D}e^{-l/(L_\mathrm{D}\!-\!1)}}{\sum_{l'=0}^{L_\mathrm{D}\!-\!1}e^{-l'/(L_\mathrm{D}\!-\!1)}}), \ \forall l \in \{0, \ldots, L_\mathrm{D}\!-\!1\}$, $[\bar{\bm{g}}_l]_m \sim \mathcal{CN}(0, \frac{\beta_\mathrm{G}e^{-l/(L_\mathrm{G}\!-\!1)}}{\sum_{l'=0}^{L_\mathrm{G}\!-\!1}e^{-l'/(L_\mathrm{G}\!-\!1)}}), \ \forall l \in \{0, \ldots, L_\mathrm{G}\!-\!1\}, \ \forall m \in \mathcal{M}$, and $[\bar{\bm{s}}_l]_m \sim \mathcal{CN}(0, \frac{\beta_\mathrm{S}e^{-l/(L_\mathrm{S}\!-\!1)}}{\sum_{l'=0}^{L_\mathrm{S}\!-\!1}e^{-l'/(L_\mathrm{S}\!-\!1)}}), \ \forall l \in \{0, \ldots, L_\mathrm{S}\!-\!1\}, \ \forall m \in \mathcal{M}$.
	We consider a noise power spectrum density of $-169~\text{dBm/Hz}$ with an additional $9~\text{dB}$ noise figure. Thus, the average receiver noise power per sub-carrier is $-169+9+10 \log_{10}(b) = -93.29~\text{dBm}$.
	We further set $\Gamma = 8.8~\text{dB}$ \cite{CioffiMulti}, $L_1 = 2.5$~\text{nH}, $L_2 = 0.7$~\text{nH}, $R = 1$~$\Omega$, and $Q=50$. All the results are averaged over $100$ independent channel realizations.
	
	\begin{figure}[t]
		\centering
		\includegraphics[width=0.36\textwidth]{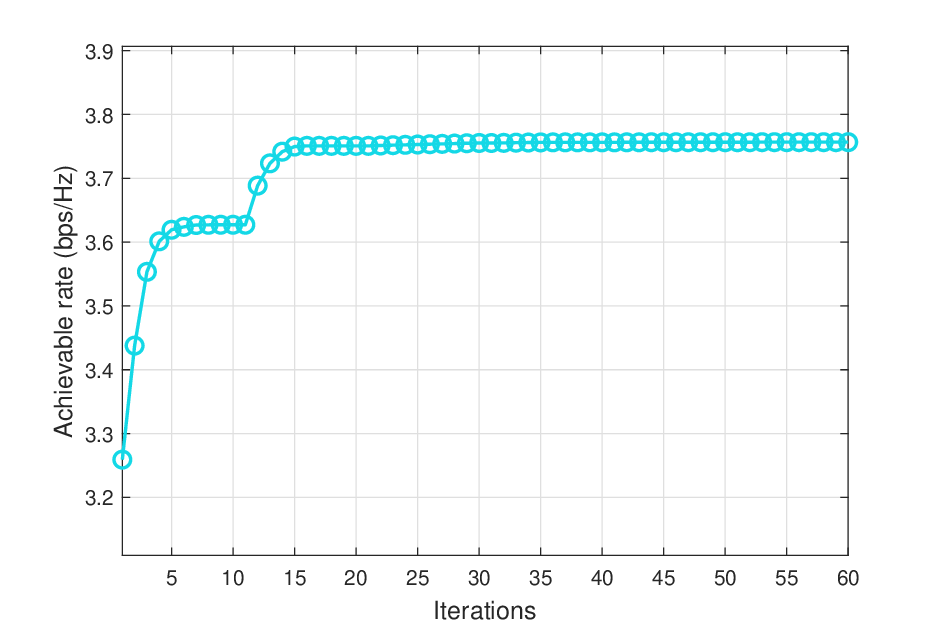} 
		\caption{Convergence behavior of the proposed algorithm.} 
		\label{fig: Iterations}
		\vspace{-0.5cm}
	\end{figure}
	
	\begin{figure}[t]
		\centering
		\includegraphics[width=0.36\textwidth]{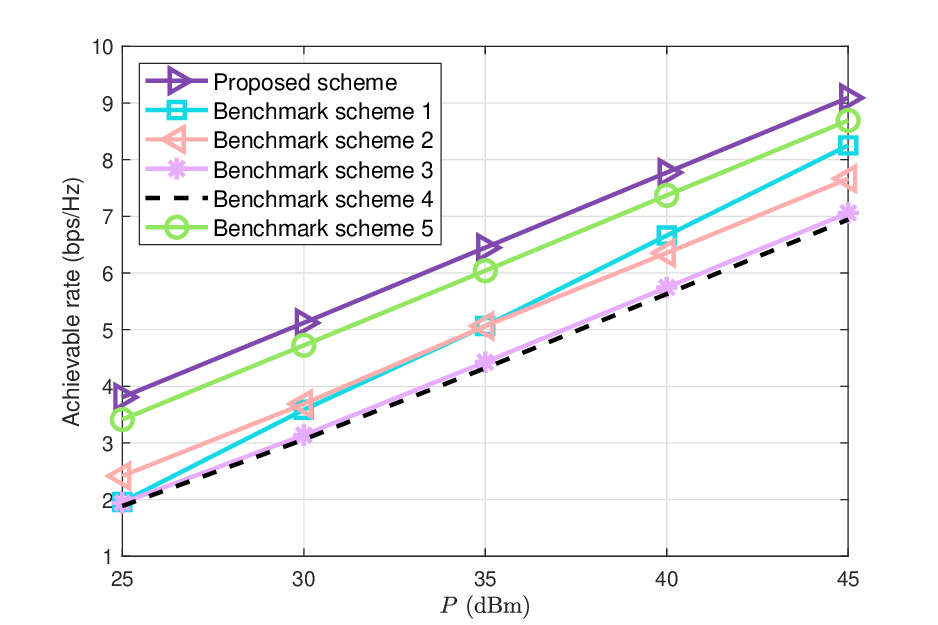} 
		\caption{Achievable rate versus transmit power.}
		\label{fig: Total_Power}
		\vspace{-0.75cm}
	\end{figure}
	
	\begin{figure}[t]
		\centering
		\includegraphics[width=0.36\textwidth]{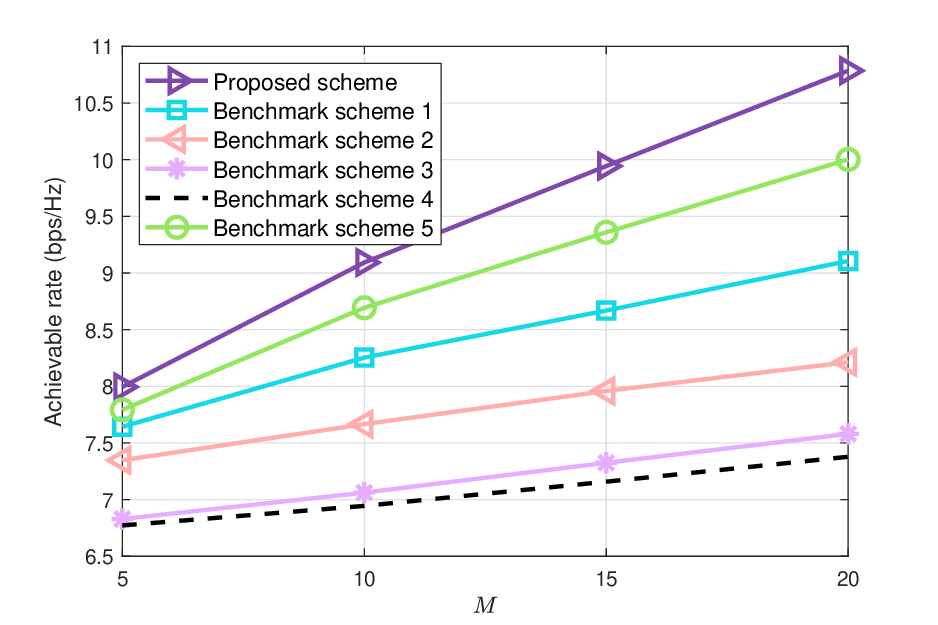} 
		\caption{Achievable rate versus number of reflecting elements.}
		\label{fig: M}
		\vspace{-0.7cm}
	\end{figure}

	First, we evaluate the convergence behavior of the proposed algorithm under $M=5$ and $P = 30~\text{dBm}$. Fig.~\ref{fig: Iterations} illustrates the achievable rate over iterations, showing monotonic convergence in the proposed algorithm, which is consistent with our discussions in Section~\ref{AO_Method}.
	Next, we compare the performance of the proposed scheme with the following benchmark schemes.
	\vspace{-0.5cm}
	\begin{itemize}
		\item \textbf{Benchmark scheme 1: Frequency-dependent BD-IRS with linear fitting method.} We consider the linear fitting method proposed in \cite{Li2024Beyond} and adopt the quasi-Newton method used in \cite{Li2024Beyond} to maximize the achievable rate.
		\item \textbf{Benchmark scheme 2: Channel power maximization.} We obtain $\{\boldsymbol{\Phi}_n\}_{n=1}^N$ and $\{p_n\}_{n=1}^N$ via our proposed initialization method in Section V-D and the WF algorithm, and then adopt the construction method in Section V-C to obtain $\bm{C}$.
		\item \textbf{Benchmark scheme 3: Frequency-independent BD-IRS.} We assume that BD-IRS admittance matrices are designed to be same for all sub-carriers, i.e., $\bm{A}_n = \bm{A}, \ \forall n \in \mathcal{N}$. Thus, BD-IRS is frequency-independent, i.e., $\boldsymbol{\Phi}_n = \boldsymbol{\Phi}, \ \forall n \in \mathcal{N}$, which is designed by extending our proposed AO-based algorithm.
		\item \textbf{Benchmark scheme 4: Conventional IRS.} We consider the conventional IRS, where $\boldsymbol{\Phi}_n$ is restricted to the diagonal structure with each diagonal element having a modulus no larger than 1. The diagonal IRS reflection matrices are designed by extending our proposed AO-based algorithm.
		\item \textbf{Benchmark scheme 5: Frequency-dependent BD-IRS with reciprocal constraint.} We add extra symmetric constraints $\boldsymbol{\Phi}_n = \boldsymbol{\Phi}_n^T, \ \forall n \in \mathcal{N}$ and $\bm{C} = \bm{C}^T$ to (P2-II-\text{eqv}') and (P3), respectively, and extend our proposed AO-based algorithm.
	\end{itemize}
	
	\looseness=-1
	In Fig.~\ref{fig: Total_Power}, we compare the achievable rate of all schemes versus the transmit power $P$ under $M = 10$. 
	It is observed that all schemes with frequency-dependent BD-IRS outperform their frequency-independent counterpart, and BD-IRS aided systems outperform the conventional IRS aided system. In addition, comparing the proposed scheme with benchmark scheme 5, it is observed that relaxing the symmetric constraints leads to a substantial performance gain.
	Moreover, the proposed scheme exhibits rate improvement compared to benchmark scheme 1, which demonstrates the superiority of the proposed algorithm. 
	When $P \leq 30~\text{dBm}$, even benchmark scheme 2 outperforms benchmark scheme 1, revealing that the linear fitting method yields significant performance loss in low signal-to-noise ratio (SNR) regimes. 
	However, as $P$ increases, the gap between benchmark scheme 1 and the proposed scheme narrows, indicating that the linear fitting method may be suitable for high SNR regimes under the considered setup.

	Furthermore, we show in Fig.~\ref{fig: M} the achievable rate versus $M$ under $P = 45~\text{dBm}$. 
	It is observed that all schemes with frequency-dependent BD-IRS outperform other schemes, and the rate gap increases with $M$. 
	Moreover, the proposed scheme achieves the best performance at all values of $M$.
	When $M$ is small, the achievable rate of benchmark scheme 1 is close to that of the proposed scheme. However, as $M$ increases, the rate gap increases, which shows that when the dimension of the fully-connected BD-IRS is moderate-to-large, the advantage of the proposed algorithm is significant.
	
	\vspace{-0.1cm}
	\section{Conclusions}
	\vspace{-0.12cm}
	\looseness=-1
	This paper studied a broadband OFDM system aided by a BD-IRS with inter-element connections. We provided an explicit model of the BD-IRS reflection matrices over different OFDM sub-carriers with respect to the tunable capacitance matrix of the BD-IRS circuit, under which we formulated the optimization problem of the sub-carrier power allocations and capacitance matrix towards OFDM communication rate maximization. Due to the coupling among the reflection matrices over different sub-carriers through the common capacitance matrix, this non-convex problem is particularly challenging to solve, for which we leveraged assorted optimization techniques to find a high-quality solution. It was verified via numerical results that our proposed design outperforms various other benchmarks.

	\bibliographystyle{IEEEtran}
	\bibliography{reference}

\begin{thebibliography}{10}
\providecommand{\url}[1]{#1}
\csname url@samestyle\endcsname
\providecommand{\newblock}{\relax}
\providecommand{\bibinfo}[2]{#2}
\providecommand{\BIBentrySTDinterwordspacing}{\spaceskip=0pt\relax}
\providecommand{\BIBentryALTinterwordstretchfactor}{4}
\providecommand{\BIBentryALTinterwordspacing}{\spaceskip=\fontdimen2\font plus
\BIBentryALTinterwordstretchfactor\fontdimen3\font minus
  \fontdimen4\font\relax}
\providecommand{\BIBforeignlanguage}[2]{{%
\expandafter\ifx\csname l@#1\endcsname\relax
\typeout{** WARNING: IEEEtran.bst: No hyphenation pattern has been}%
\typeout{** loaded for the language `#1'. Using the pattern for}%
\typeout{** the default language instead.}%
\else
\language=\csname l@#1\endcsname
\fi
#2}}
\providecommand{\BIBdecl}{\relax}
\BIBdecl

\bibitem{Shen2022Modeling}
S.~Shen, B.~Clerckx, and R.~Murch, ``Modeling and architecture design of
  reconfigurable intelligent surfaces using scattering parameter network
  analysis,'' \emph{IEEE Trans. Wireless Commun.}, vol.~21, no.~2, pp.
  1229--1243, Feb. 2022.

\bibitem{11054049}
S.~Zheng and S.~Zhang, ``Beyond diagonal intelligent reflecting surface aided
  integrated sensing and communication,'' \emph{IEEE Trans. Cogn. Commun.
  Netw.}, early access, doi: 10.1109/TCCN.2025.3583699.

\bibitem{Yang2020Intelligent}
Y.~Yang, B.~Zheng, S.~Zhang, and R.~Zhang, ``Intelligent reflecting surface
  meets {OFDM}: Protocol design and rate maximization,'' \emph{IEEE Trans.
  Commun.}, vol.~68, no.~7, pp. 4522--4535, Jul. 2020.

\bibitem{Yang2020IRS}
Y.~Yang, S.~Zhang, and R.~Zhang, ``{IRS}-enhanced {OFDMA}: Joint resource
  allocation and passive beamforming optimization,'' \emph{IEEE Wireless
  Commun. Lett.}, vol.~9, no.~6, pp. 760--764, Jun. 2020.

\bibitem{Zhang2020Capacity}
S.~Zhang and R.~Zhang, ``Capacity characterization for intelligent reflecting
  surface aided {MIMO} communication,'' \emph{IEEE J. Sel. Areas Commun.},
  vol.~38, no.~8, pp. 1823--1838, Aug. 2020.

\bibitem{Li2021Intelligent}
H.~Li, W.~Cai, Y.~Liu, M.~Li, Q.~Liu, and Q.~Wu, ``Intelligent reflecting
  surface enhanced wideband {MIMO-OFDM} communications: From practical model to
  reflection optimization,'' \emph{IEEE Trans. Commun.}, vol.~69, no.~7, pp.
  4807--4820, Jul. 2021.

\bibitem{Demir2024Wideband}
{\"O}.~T. Demir and E.~Björnson, ``Wideband channel capacity maximization with
  beyond diagonal {RIS} reflection matrices,'' \emph{IEEE Wireless Commun.
  Lett.}, vol.~13, no.~10, pp. 2687--2691, Oct. 2024.

\bibitem{Soleymani2024Maximizing}
M.~Soleymani, I.~Santamaria, A.~Sezgin, and E.~Jorswieck, ``Maximizing spectral
  and energy efficiency in multi-user {MIMO OFDM} systems with {RIS} and
  hardware impairment,'' [Online]. Available: https://arxiv.org/abs/2401.11921.

\bibitem{Katsanos2024Multi}
K.~D. Katsanos, P.~D. Lorenzo, and G.~C. Alexandropoulos,
  ``Multi-{RIS}-empowered multiple access: A distributed sum-rate maximization
  approach,'' \emph{IEEE J. Sel. Top. Signal Process.}, vol.~18, no.~7, pp.
  1324--1338, Oct. 2024.

\bibitem{Sena2024Beyond}
A.~S.~de Sena, M.~Rasti, N.~H.~Mahmood, and M.~Latva-aho, ``Beyond diagonal
  {RIS} for multi-band multi-cell {MIMO} networks: A practical
  frequency-dependent model and performance analysis,'' \emph{IEEE Trans.
  Wireless Commun.}, vol.~24, no.~1, pp. 749--766, Jan. 2025.

\bibitem{Li2024Beyond}
H.~Li, M.~Nerini, S.~Shen, and B.~Clerckx, ``Beyond diagonal reconfigurable
  intelligent surfaces in wideband {OFDM} communications: Circuit-based
  modeling and optimization,'' \emph{IEEE Trans. Wireless Commun.}, vol.~24,
  no.~4, pp. 3623--3636, Apr. 2025.

\bibitem{CioffiMulti}
\BIBentryALTinterwordspacing
J.~M. Cioffi, ``Multi-channel modulation,'' Stanford Univ., Stanford, CA, USA,
  Tech. Rep. Accessed: Jul. 2019. [Online]. Available:
  \url{https://cioffi-group.stanford.edu/doc/book/chap4.pdf}
\BIBentrySTDinterwordspacing

\bibitem{Nerini2024A}
M.~Nerini, S.~Shen, H.~Li, M.~Di~Renzo, and B.~Clerckx, ``A universal framework
  for multiport network analysis of reconfigurable intelligent surfaces,''
  \emph{IEEE Trans. Wireless Commun.}, vol.~23, no.~10, pp. 14\,575--14\,590,
  Oct. 2024.

\bibitem{Pozar2009Microwave}
D.~M. Pozar, \emph{Microwave Engineering}.\hskip 1em plus 0.5em minus
  0.4em\relax Hoboken, NJ, USA: Wiley, 2009.

\bibitem{10993452}
H.~Li and B.~Clerckx, ``Non-reciprocal beyond diagonal {RIS}: Multiport network
  models and performance benefits in full-duplex systems,'' \emph{IEEE Trans.
  Commun.}, early access, doi: 10.1109/TCOMM.2025.3568222.

\bibitem{goldsmith2005wireless}
A.~Goldsmith, \emph{Wireless Communications}.\hskip 1em plus 0.5em minus
  0.4em\relax Cambridge, U.K.: Cambridge Univ. Press, 2005.

\bibitem{marks1978A}
B.~R. Marks and G.~P. Wright, ``A general inner approximation algorithm for
  nonconvex mathematical programs,'' \emph{Oper. Res.}, vol.~26, no.~4, pp.
  681--683, Jul. 1978.

\bibitem{cvx}
\BIBentryALTinterwordspacing
M.~Grant and S.~Boyd. {(Jan. 2020). \textit{CVX: MATLAB Software for
  Disciplined Convex Programming}}. [Online]. Available:
  \url{http://cvxr.com/cvx/}
\BIBentrySTDinterwordspacing

\bibitem{Hong2016Unified}
M.~Hong, M.~Razaviyayn, Z.-Q. Luo, and J.-S. Pang, ``A unified algorithmic
  framework for block-structured optimization involving big data: With
  applications in machine learning and signal processing,'' \emph{IEEE Signal
  Process. Mag.}, vol.~33, no.~1, pp. 57--77, Jan. 2016.

\bibitem{Luo2010Semidefinite}
Z.-Q. Luo, W.-K. Ma, A.~M.-C. So, Y.~Ye, and S.~Zhang, ``Semidefinite
  relaxation of quadratic optimization problems,'' \emph{IEEE Signal Process.
  Mag.}, vol.~27, no.~3, pp. 20--34, May 2010.

\bibitem{Zhang2018Constant}
S.~Zhang, R.~Zhang, and T.~J. Lim, ``Constant envelope precoding for {MIMO}
  systems,'' \emph{IEEE Trans. Commun.}, vol.~66, no.~1, pp. 149--162, Jan.
  2018.

\end{thebibliography}
	
\end{document}